\documentclass{article}
\usepackage[dvips]{graphicx}
\def\be{\begin{equation}}
\def\ee{\end{equation}}
\def\bea{\begin{eqnarray}}
\def\eea{\end{eqnarray}}

         \def\la{\label}
         \def\rf{\ref}

         \def\se{\section}
         
         \def\o{\over}
         \def\a{\alpha}
         \def\b{\beta}
         
         \def\bbz{{\rm Z\mkern-8mu Z}}
\begin{document}
\begin{titlepage}
\vspace{1cm}
\begin{center} 

{\Large \bf  A Multi-Species Asymmetric Exclusion Model with an Impurity}\\
\vspace{1cm}
\centerline {\bf \mbox{Farhad H Jafarpour} \footnote {e-mail:farhad@mpipks-dresden.mpg.de}}
{\vspace{1cm}
Max-Planck-Institut f\"ur Physik Complexer Systeme, \\
N\"othnitzer Str. 38, D-01187 Dresden, Germany}\\
\end{center}

\vskip 2cm

\begin{abstract}
A multi-species generalization of the Asymmetric Simple Exclusion Process (ASEP) has been considered in the 
presence of a single impurity on a ring. The model describes particles hopping in one direction with stochastic dynamics  
and hard core exclusion condition. The ordinary particles hop forward with their characteristic hopping rates and fast particles can 
overtake slow ones with a relative rate. The impurity, which is the slowest particle in the ensemble of particles on the ring,
hops in the same direction of the ordinary particles with its intrinsic hopping rate and can be overtaken by ordinary particles 
with a rate which is not necessarily a relative rate.
We will show that the phase diagram of the model can be obtained exactly. It turns out that the phase structure of the 
model depends on the density distribution function of the ordinary particles on the ring so that it can have either four phases 
or only one. 
The mean speed of impurity and also the total current of the ordinary particles are explicitly calculated in each 
phase. Using Monte Carlo simulation, the density profile of the ordinary particles is also obtained. The simulation data confirm 
all of the analytical calculations.
\end{abstract}
\end{titlepage}

\section{Introduction}
One of the simplest models of driven diffusive systems is the {\it Asymmetric Simple Exclusion Process} (ASEP) \cite{sp,sz,l,sph}. 
The ASEP consists of
particles on a one-dimensional lattice $\bbz$ attempting to jump to an adjacent site, choosing the site to their right
with rate $p$, and to their left with rate $q$, with the attempt succeeding if the target site is not already occupied.   
This model can describe the Burgers equation on a microscopic scale \cite{kr1}. However the ASEP is not only interesting because of
its relation to the Burgers equation but also it has many applications in other fields such as growth models, traffic and kinetics 
of portein synthesis \cite{kd,kr2,chsa,mac}. 
During last decade the ASEP has been studied extensively both with open boundaries and periodic boundary
condition. It is known that this simple model shows a variety of interesting phenomena such as boundary-induced phase transitions and 
shock profiles \cite{schr,d,dehp,de}.\\
In the open boundary case the particles are injected and also extracted from both ends of the chain so that the density 
of particles on the chain might vary. In this case the density of the particles on the chain can be controlled by some boundary
parameters, which in fact control the input and output rates. Depending on the value of these parameters, 
the phase diagram of the model can possesses three different phases: a high-density phase (HD), 
a low-density phase (LD) and a maximal-current phase (MC). It is known that the density profile of the particles has a shock structure
\footnote{The "shock structure" means a sharp discountinuty between a region of high density of particles and a region of low density.}
on the coexistance line between LD and HD phase where a first order phase transtion takes 
place. A second order phase transition also exists from LD and HD phases to the MC phase \cite{sand,esri,sas1}.
This phase structure seems to be generic for all those models in which the fundamental digram
(the current of particles as a function of the mean density of them) has a single local maximum. 
The ASEP with periodic boundary condition, where the total number of particles on the chain is a conserved quantity,
in the presence of a second class particle has also been studied in two different cases \cite{djls,m,farhad,sasa2,lpk}.
In the first case the second class particle (which is called the impurity) hops only to the right. The ordinary particles hop 
both to the left and right and can also overtake the impurity from the left. The phase diagram of the model in comparison to the open 
boundary problem is rather complicated. Again the system presents a shock in one of the phases.   
In the second case the impurity hops only to the left. The ordinary particles and the impurity may exchange their positions with a
finite rate. This model has only been studied for $q=0$, and the case $q\neq 0$ is still an open probelm.
It has been shown that the phase diagram of the model for $q=0$ has only two different phases: A shock phase and   
a low-density phase \cite{lpk}. In the absence of the second class particle the stationary state is well known: all configurations 
are equally likely.\\
Recently, a multi-species generalization of the ASEP which consists of $p$ different species of particles hopping in one 
direction has been introduced \cite{k1}. 
Since the number of species is quite arbitrary, one can take the infinite species limit. In this limit 
the hopping rate of each species of particles $v$ is taken from a continuous distribution $\sigma(v)$. In this model the overtaking
between particles is allowed so that
the fast particles can overtake the slow ones. Particles of different kinds can also be inserted
and extracted from the system. Using a Matrix Product formalism (first introduced in \cite{dehp})  
the phase diagram of this model in the stationary state has been obtained \cite{k3}. It turns out that this multi-species generalization
has a richer phase structure (in comparison to the usual ASEP) depending on the distribution function $\sigma(v)$ so that it can exist
either in a three-phase
or a two-phase regime. The phase diagram of the model in the three-phase regime is nearly similar to the phase diagram of the ASEP
with open boundaries. However, in the two-phase regime the high density region disappears.\\
In this paper we study the multi-species ASEP in the presence of an impurity on a ring. We will consider the case in which the
impurity moves in the same direction of the ordinary particles and can be overtaken by them. In the periodic boundary condition case, 
instead of the distribution function $\sigma(v)$ which gives the probability of finding a particle with hopping rate $v$ {\em before 
entering the system}, we introduce $\rho(v)$ which gives the density of the ordinary particles with hopping rate $v$ on the ring.
This quantity can also give the probability of finding an ordinary particle with intrinsic hopping rate $v$ on the ring at any time.
We will show that the phase structure of this model depends on the density distribution function of
normal particles $\rho(v)$ on the ring so that it can be in two different regimes. In the first regime the phase diagram is
rather similar to the phase diagram of the one-species ASEP with an impurity on a ring (the first case) \cite{m} and it has four 
different phases. There exists a region in which the
total density profile of the ordinary particles has a shock structure. In the one-species limit the speed of impurity in the
shock phase is independent of the total density of the ordinary particles on the ring. However, for our multi-species model the speed of 
impurity in this phase, which is in fact equal to the speed of the shock, depends on $\rho(v)$.
In the second regime, we may see two different behaviours. If the speed of impurity is smaller than the hopping rate of the slowest 
ordinary particle on the ring, then the phase diagram will contain only one phase. However, if we allow the impurity have
a hopping rate greater than the hopping rate of the slowest ordinary particle, then a new phase will appear so that the phase diagram
in this regime will contain two different phases. In order to check the analytical results we have done the Monte Carlo simulations.
It turns out that the simulation data confirms all of the analytical results. Our analytical approach does not let us calculate the
density profile of the ordinary particles on the ring; however, we will try to obtain the density profile of the ordinary particles on the
ring using the Monte Carlo simulations. The multi-species ASEP has also been studied on a ring in the absence of the impurity \cite{k2}.
It is known that the stationary state of this model is uncorrelated. We will show that on an specific line in the parameters space
the impurity behaves like an ordinary particle so that on this line the same results of \cite{k2} can be derived.\\       
This paper is organized as follows. In section 2, we will define the model and  
obtain its generating function for the grand canonical partition function on the ring. 
Analyzing the singularities of this generating function lets us find the phase structure and also calculate the mean value of two 
important 
physical quantities, i.e. the mean speed of the impurity and the total current of the ordinary particles in the stationary state.
In section 3, we will study two different examples to check the validity of our calculations: 
First we will take the one-species limit and show that all of our results
converge on those obtained in \cite{m}. In the second example we will introduce a family of density distribution functions, and 
using Monte Carlo simulation show that all of the analytical results obtained for this family of distribution are correct.
In the last section we will present the conclusions.
\section{Exact Stationary State Using the Matrix Product Formalism}
\subsection{Definition of the Model}
Our model consists of $p$ species of first class particles (we refer to them as the ordinary particles)
and a single second class particle (we refer to it as the impurity) on a chain of length $L+1$ with 
periodic boundary condition. We assume that the number of particles on the ring is a conserved quantity.
Each species of particles has its own intrinsic hopping rate. Fast ordinary particles overtake the slow ones
with their relative hopping rate. The impurity has the smallest hopping rate in the ensemble of the particles on the ring and thus, can 
be overtaken by ordinary particles of any kind with a rate which is not necessarily a relative rate.  
In this paper, following \cite{k3}, we take the infinite species limit, since the number of species is an arbitrary number. In this limit the 
hopping rate of the ordinary particles $v$ is a continuous parameter and the density of particles in each species has also been taken 
from a continuous density distribution $\rho(v)$. The function $\rho(v)$ is defined in the interval $[v_1,+\infty($
in which $v_1$ is the smallest hopping rate in the ensemble of the ordinary particles. 
Denoting an ordinary particle with the intrinsic hopping rate $v$ by $D(v)$, 
the impurity by $A$ and an empty site by $E$, during the infinitesimal time step $dt$ any bound ($i$,$i+1$) (with $1\leq i \leq L+1$) 
evolves as follows:
\bea
D(v) E     &\longrightarrow& E D(v) \ \ \ \ \ \ \ \mbox{with the rate} \ \ v \nonumber \\
D(v) D(v') &\longrightarrow& D(v') D(v) \ \       \mbox{with the rate} \ \ v-v' \ \ , \ \ v>v' \\
D(v) A     &\longrightarrow& A D(v) \ \ \ \ \ \ \ \mbox{with the rate} \ \ v+\beta-1 \nonumber \\
A E        &\longrightarrow& E A  \ \ \ \ \ \ \ \ \ \ \ \mbox{with the rate} \ \ \alpha \nonumber
\eea
in which $\alpha$ and $\beta$ are positive parameters. We specify each ordinary particle by its intrinsic hopping rate hereafter.
According to the above process, particles hop to their rightmost
site with their intrinsic hopping rate, provided that it is empty. If an ordinary particle of kind $v$ encounters 
a site occupied by an ordinary partice of kind $v'$, they might exchange their positions with rate $v-v'$ provided that $v>v'$.
An ordinary particle of kind $v$ can also overtake the impurity with rate $v+\beta-1$. For this rate to be positive, we require 
that $\beta\geq 1-v_1$. It can be seen that, as far as $\alpha+\beta\neq 1$
the impurity behaves quite differently from the ordinary particles. For $\alpha+\beta =1$, since $\alpha<v_1$, the impurity 
behaves like an ordinary particle with the smallest hopping rate. On this line, all the ordinary particles overtake the impurity with 
a relative rate. It is known that the steady state of the system in this case is completely uncorrelated and all the 
configurations have the same probability of occuring \cite{k2}. Also, the special case $\rho(v)=\rho\delta(1-v)$ has already been studied
in \cite{m}. In this case all of the ordinary particles have the same hopping rate (one-species limit) so that 
no overtaking between them takes 
place. In order to study the stationary state properties of this model, we will introduce the necessary mathematical tools in the 
following section. 
\subsection{Algebraic Preliminaries}
Among different methods the Matrix Product formalism \footnote{This formalism is also known as the Matrix Product Ansatz (MPA) 
in related literature.} has proved to be one of the most useful for obtaining the stationary state properties of reaction-diffusion
models. According to this formalism the stationary probability $P(\{C\})$ of any configuration $\{C\}$ can be 
written as a trace of the product of non-commuting operators \cite{ks}. In our model, for the sake of 
simplicity, we will work in the relative frame of the impurity i.e. using the translational invariance of the system we always 
keep the impurity at site $L+1$. We will also use a grand canonical ensemble, in which the number of ordinary particles in each species 
fluctuates arround a mean value. However, the density of each species should be fixed by its fugacity. By using the MPA the stationary 
probability of any configuration, $\{C\}$ can be written as follows:
\be
\la{p}
P(\{C\})=\frac{1}{Z_L}Tr(\prod_{i=1}^{L+1}X_i)
\ee
in which $X_i=E$ ($1\leq i \leq L$), if site $i$ is empty, and $X_i=z(v)D(v)$ ($1\leq i\leq L$) 
if it is occupied by an ordinary particle of kind $v$. Since the site $L+1$ is always occupied by the impurity, we take $X_{L+1}=A$.
The quantity $z(v)$ is the fugacity of the ordinary particles of kind $v$ which should be fixed using:
\be
\la{mean}
\rho(v)=lim_{L\rightarrow\infty}\frac{z(v)}{L}\frac{\delta}{\delta z(v)} \ln Z_{L}.
\ee
We have not introduced the fugacity of holes since the total number of available sites $L+1$ is fixed.
The total density of the ordinary particles is also specified by $\int_{v_1}\rho(v)dv=\rho$. The normalization factor
$Z_{L}$ in the denominator of (\rf{p}) plays the role analogous with the grand canonical partition function in equilibrium statistical
mechanics, and can be calculated using the fact that $\sum_{C}P(\{C\})=1$. Thus one finds:
\be
\la{parti}
Z_L=Tr(C^L A) \ \ , \ \ C=E+\int_{v_1} z(v)D(v)dv. 
\ee
As we will see shortly, the mean value of the physical quantities such as the speed of impurity and the current of each species of ordinary
particles can be written in terms of $Z_L$. The operators $E$, $D(v)$'s and $A$ then satisfy the following relations:
\bea
\la{alg}
\begin{array}{ll}
D(v) E = {1\over v} D(v) + E  &   \\
D(v') D(v) =  \frac{1}{v - v'}(v D(v') - v'D(v)),  &  v'> v \\ 
D(v) A = {v\over {v+\b-1}} A & \\
AE = {1\over \a}A & 
\end{array}
\eea
Following \cite{efgm} we write $A$ as $\vert V\rangle\langle W\vert$ which simplifies the above algebra to the one in
\cite{k3} for the study of the multi-species ASEP with open boundaries. \\
In the following we will show how one can calculate the mean value of physical quantities in the stationary state using the 
MPA. For example, the speed of impurity in the reference frame of the lattice can be written as:
$$
V=\alpha\frac{Tr(C^{L-1}AE)}{Tr(C^LA)}-\int_{v_1}(v+\beta-1)\frac{Tr(C^{L-1}z(v)D(v)A)}{Tr(C^LA)}dv.
$$
Using the algebra of (\rf{alg}) one can easily simplify the above expression to find:
\be
\la{V}
V=(1-\int_{v_1}vz(v)dv)\frac{Tr(C^{L-1}A)}{Tr(C^LA)}=(1-\int_{v_1}vz(v)dv)\frac{Z_{L-1}}{Z_{L}}.
\ee
One can also find a simple expression for the total current of the ordinary particles on the ring $J_{total}$ in the stationary state 
as follows: We specify
each configuration of the system by an $L+1$-tuple ($\tau_1,\cdots,\tau_{L+1}$) where $\tau_i=v$ if site $i$ is
occupied with an ordinary particle of kind $v$, $\tau_i=a$ if it is occupied by the impurity, and $\tau_i=e$ if it is empty.
Now one can write the current of the ordinary particles of kind $v$, $J(v)$, in the stationary state as follows:
\bea
\begin{array}{c}
J(v)=vP(\tau_k=v,\tau_{k+1}=e)+\int_{v'<v}(v-v')P(\tau_k=v,\tau_{k+1}=v')dv' \nonumber \\
-\int_{v'>v}(v'-v)P(\tau_k=v',\tau_{k+1}=v)dv'+(v+\beta-1)P(\tau_k=v,\tau_{k+1}=a)  \nonumber
\end{array}
\eea
where $P(\tau_{k}=\cdots,\tau_{k+1}=\cdots)$ is the probability of finding the system in configuration $\{ C \}$ provided that 
the sites $k$ and $k+1$ are occupied by particles or holes. The above relation in turn can be written in terms of the conditional
probability $P(\tau_{k}=\cdots,\tau_{k+1}=\cdots\vert \tau_l=\cdots)$ given by (\rf{p}) which has the same definition as 
$P(\tau_{k}=\cdots,\tau_{k+1}=\cdots)$ provided that the impurity is at site $l$:
\bea
\begin{array}{c}
J(v)=v\sum_{l=1}^{'L-1}P(\tau_k=v,\tau_{k+1}=e\vert\tau_l=a)P(\tau_l=a)+ \nonumber \\
\sum_{l=1}^{'L-1}\int_{v'<v}(v-v')P(\tau_k=v,\tau_{k+1}=v'\vert\tau_l=a)P(\tau_l=a)dv' \nonumber \\
-\sum_{l=1}^{L-1}\int_{v'>v}(v'-v)P(\tau_k=v',\tau_{k+1}=v\vert\tau_l=a)P(\tau_l=a)dv'
+(v+\beta-1)P(\tau_k=v,\tau_{k+1}=a). \nonumber
\end{array}
\eea
where the prime over a sum means that the terms $l=k,k+1$ should be excluded.
Since the system in the stationary state is translationally invariant, the impurity can be on any of the $L+1$ sites with the 
same probability. Now using the algebra (\rf{alg}) we obtain:
$$
P(\tau_k=v,\tau_{k+1}=a)=P(\tau_k=v\vert\tau_{k+1}=a)P(\tau_{k+1}=a)=\frac{1}{L+1}\frac{vz(v)}{v+\beta-1}
\frac{Tr(C^{L-1}A)}{Tr(C^{L}A)}
$$
which after calculations, results in the following expression for the current of the ordinary particles of kind $v$:
\bea
\begin{array}{c}
J(v)=vz(v)\frac{L}{L+1}\frac{Tr(C^{L-1}A)}{Tr(C^{L}A)}+ 
\frac{1}{L+1}\sum_{k=1}^{L-1}\frac{Tr(C^{k-1}z(v)D(v)C^{L-k-1}A)}{Tr(C^LA)}-\nonumber  \\
vz(v)\frac{1}{L+1}\sum_{k=1}^{L-1}\int_{v_1}\frac{Tr(C^{k-1}z(v')D(v')C^{L-k-1}A)}{Tr(C^LA)}dv'+ \nonumber \\
\sum_{k=1}^{L-1}\int_{v'<v}\frac{z(v)z(v')(v-v')}{L+1}\frac{Tr(C^{k-1}D(v)D(v')C^{L-k-1}A)}{Tr(C^LA)}dv'-\nonumber  \\
\sum_{k=1}^{L-1}\int_{v'>v}\frac{z(v)z(v')(v'-v)}{L+1}\frac{Tr(C^{k-1}D(v')D(v)C^{L-k-1}A)}{Tr(C^LA)}dv'.
\end{array}
\eea
Using the definition of the total density of the ordinary particles of kind $v$ (\rf{mean}):
$$
\rho(v)=\frac{1}{L}\sum_{k=0}^{L-1}\frac{Tr(C^{k}z(v)D(v)C^{L-k-1}A)}{Tr(C^L A)}
$$
and also the algebra of (\rf{alg}) one can find the following simple expression for $J(v)$:
\bea
\begin{array}{c}
J(v)=[\frac{L-1}{L+1}\rho(v)+\frac{L}{L+1}z(v)v-\frac{L-1}{L+1}z(v)v\rho]\frac{Tr(C^{L-1}A)}{Tr(C^LA)}+ \nonumber \\
\frac{L-1}{L+1}[(\int_{v'<v}-\int_{v'>v})(z(v)v\rho(v')-z(v')v'\rho(v))dv'] \frac{Tr(C^{L-1}A)}{Tr(C^LA)}.\nonumber
\end{array}
\eea
In the thermodynamic limit $L\longrightarrow \infty$ we find:
$$
J(v)=[\rho(v)+(1-\rho)z(v)v+(\int_{v'<v}-\int_{v'>v})(z(v)v\rho(v')-z(v')v'\rho(v))dv'] \frac{Z_{L-1}}{Z_L}.
$$   
In order to calculate the total current to the ordinary particles $J_{total}$, one should integrate the above expression over 
all species:
\bea
\begin{array}{c}
\la{J}
J_{total}=\int_{v_1} J(v)dv= \\
\rho V +(\int_{v_1}vz(v)dv)\frac{Z_{L-1}}{Z_L}=\rho V + (\frac{Z_{L-1}}{Z_L}-V).
\end{array}
\eea
As can be seen both $J_{total}$ and $V$ are functions of $\frac{Z_{L-1}}{Z_L}$ and also independent of the position on the ring 
as one should expect in the steady state.
In principle, one can calculate these quantities using the non-trivial (non-commuting) representations of the algebra (\rf{alg}).
It has been shown that all non-commuting representations of (\rf{alg}) are infinite dimensional \cite{k1}. However, 
the author of \cite{k1} has shown that a coherent basis exists which using it one can compute the generating function 
for the canonical partition function of this model with open boundaries \cite{k3}. 
Finding the singularities of this generating function lets us obtain not only the exact phase diagram but also the mean values of 
the physical quantities.
The generating function for the grand canonical partition function of our model which is defined as:
\be
\la{gener}
f({\bf z};s;\alpha,\beta)=\sum_{L=0}^{\infty}s^LTr(C^LA)=\sum_{L=0}^{\infty}s^L \langle W \vert C^L \vert V \rangle=
\sum_{L=0}^{\infty}s^L Z_L .
\ee
can also be calculated exactly using the same coherent basis. 
In fact (\rf{gener}) has already been calculated to obtain the mean density of the ordinary
particles in the middle of chain in the open boundary problem (see equation (64) of \cite{k3}) and is given by the following exact 
expression: 
\be
\la{gener2}
f({\bf z};s;\alpha,\beta)=\frac{\frac{\alpha B({\bf z};s)}{\alpha-B({\bf z};s)}-
\frac{\alpha (1-\beta)}{\alpha+\beta-1}}{(1-\beta)[\frac{s}{S({\bf z};1-\beta)}-1]}
\ee
in which:
\be
\la{S}
S({\bf z};u)=\frac{1}{\frac{1}{u}+\int_{v_1}\frac{z(v)}{1-\frac{u}{v}}dv} \ \ , \ \ u\in[0,v_1].
\ee
In (\rf{gener2}) the function $B({\bf z};s)$ is the inverse of $S({\bf z};u)$ in the interval $u\in[0,u_m]$ which tends to zero as 
$s\rightarrow 0$ and also has the following properties:
\be
\cases{S[{\bf z};B(s)]=s,&$0\leq s\leq s_m$\cr
       B[{\bf z};S(u)]=u,&$0\leq u\leq u_m.$\cr}
\ee
In figure 1 we have plotted $S({\bf z};u)$ and also $B({\bf z};s)$ as a function of $u$ and $s$ respectively.
One can easily see from (\rf{S}) that for $u\in[0,v_1]$, $\frac{d^2}{du^2}S^{-1}$ is always positive so $S^{-1}$ is a 
concave function in that region. On the other hand 
since $z(v)$ is a positive real number, (\rf{S}) is a positive function of $u$ for $u\in[0,v_1]$ so it can only have one maximum $u_m$ 
in this interval. 
\begin{figure}
\centering
\includegraphics[height=10cm,angle=270]{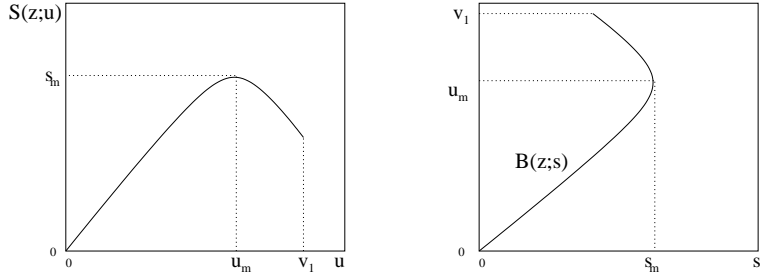}
\caption{\small{Plot of $S({\bf z};u)$ (Left) as a function of $u$ in the interval $u\in[0,v_1]$ and also $B({\bf z};s)$ (Right) 
the inverse function of $S({\bf z};u)$ in the interval $s\in[0,s_m]$ as a function of $s$. }}
\end{figure} 
The existence of the local maximum depends on the sign of
\be
\la{cond}
\frac{d}{du}S\Big\vert_{u=v_1}\propto\frac{1}{{v_1}^2}-\int_{v_1}\frac{vz(v)}{(v-v_1)^2}dv.
\ee
Because of the relation (\rf{mean}), the above quantity in turn is a function of the density distribution function of the ordinary 
particles on the ring $\rho(v)$. The generating function (\rf{gener2}) is restricted to the region $\alpha+\beta>1$ so in order to have
access to the whole parameters space we will have to introduce an analytical continuation. \\
In the next section we will obtain both the phase diagram and also the values of $V$ and $J_{total}$ 
by analyzing the singularities of (\rf{gener2}).
\subsection{Exact Results for the Physical Quantities and the Phase Diagram}
As we mentioned, the phase structure of the model and also 
the mean value of the physical quantities are determined by the singularities of (\rf{gener2}). The number of
singularities of (\rf{gener2}) depends on whether the local maximum exists or not. 
If (\rf{cond}) is negative, then the local maximum exsists, and (\rf{gener2}) has four singularities depending 
on the values of $\alpha$ and $\beta$:
\bea
\la{1sing1}
&s_\alpha=S({\bf z};\alpha)& \ \ \mbox{for} \ \ \alpha<u_m \ , \ S({\bf z};\alpha)<S({\bf z};1-\beta) \\
\la{1sing2}
&s_\beta=S({\bf z};1-\beta)&  \ \ \mbox{for} \ \ \beta<1-u_m \ , \ S({\bf z};1-\beta)<S({\bf z};\alpha) \\
\la{1sing3}
&s_m=S({\bf z};u_m)&     \ \ \mbox{for} \ \ \alpha>u_m \ , \ \beta>1-u_m \\
\la{1sing4}
&s_0=s_\alpha=s_\beta &     \ \ \mbox{for} \ \ \alpha<u_m \ , \ \beta<1-u_m.
\eea
The first (second) singularity arises from the disappearance of $\alpha-B({\bf z};s_\alpha)$ ($\frac{s_\beta}{S({\bf z};1-\beta)}-1$) 
in the numerator (denominator) of (\rf{gener2}). 
The third singularity occurs at $s_m$ where $S({\bf z};u)$ has a local maximum at $u_m$, and so $B$ becomes singular. In order to 
calculate $u_m$ in terms of the density distribution function $\rho(v)$, we note that the convergence radius of the formal series 
(\rf{gener}) can be written as: 
\be
\la{R}
R({\bf z})=\lim_{L\longrightarrow \infty}\Big(\langle W \vert C^L\vert V \rangle \Big)^{-1\o L}
\ee
so the mean density of the ordinary particles of kind $v$ in each phase (\rf{mean}) can be written in terms of the convergence radius 
$R({\bf z})$:
\be
\la{mean2}
\rho(v)=z(v)\frac{\delta}{\delta z(v)} \ln \frac{1}{R({\bf z})}.
\ee
On the other hand, the convergence radius of (\rf{gener}) is the absolute value of its nearest singularity to the origin;
therefore, from (\rf{S}), (\rf{1sing3}) and (\rf{mean2}) we arrive at the following expression which gives $u_m$ only as a function 
of $\rho(v)$:
\be
\la{bm2}
\int_{v_1}\frac{v \rho(v)}{v-u_m}dv=1.
\ee
The last singularity takes place whenever $s_\alpha=s_\beta$. Using (\rf{S}) one finds the following condition on $z(v)$, 
$\alpha$ and $\beta$ in this phase:
\be
\la{I}
\int_{v_1}\frac{vz(v)dv}{(v-\alpha)(v+\beta-1)}=\frac{1}{\alpha(1-\beta)}.
\ee
The above condition, which is the consequence of using the grand canonical ensemble, means that only two of the mentioned quantities 
can be chosen freely. We consider two different cases: First, we assume that $\beta$ is a dependent variable. 
By using (\rf{S}) and (\rf{mean2}) we obtain:
\be
\la{II}
\rho(v)=\alpha(1-\rho)\frac{vz(v)}{v-\alpha}.
\ee
Substituting (\rf{II}) in (\rf{I}) gives:
\be
\la{A}
\int_{v_1}\frac{v\rho(v)dv}{v+\beta-1}=1.
\ee
Now comparing (\rf{bm2}) and (\rf{A}) this specifies one of the boundaries of this phase i.e. $u_m=1-\beta$. Similarly, taking 
$\alpha$ as dependent variable, one finds $u_m=\alpha$ which gives the other boundary of this phase. 
In the open boundary problem $s_0$ determines a line in the parameter space. However, in our model it specifies a region in which, as
simulation results show, a shock profile evolves in the density profile of the ordinary particles on the ring. 
In order to determine the boundaries of different phases one should note that using (\rf{S}) and
(\rf{mean2}) the condition $s_{\alpha}<s_{\beta}$ in (\rf{1sing1}) can be written as $\beta>1-u_m$. Similarly one can see that the
condition $s_{\beta}<s_{\alpha}$ in (\rf{1sing2}) can be written as $\alpha>1-u_m$. Unforunately, we have not been able to find a simple
expression for the total density profile of the ordinary particles on the ring defined as:
\be
\la{total}
\rho_{total}(i)=\int_{v_1}\rho(v,i)dv=\int_{v_1}\frac{Tr(C^{i-1}z(v)D(v)C^{L-i}A)}{Tr(C^LA)}dv
\ee
where $\rho(v,i)$ is the density of the ordinary particles of kind $v$ at site $i$.
However, we can find the exact values of $\rho_{total}(i)$ exactly in front of the impurity $\rho_{total}(1)$, and also behind it
$\rho_{total}(L)$. For $\rho_{total}(1)$ we obtain:
\be
\rho_{total}(1)=\int_{v_1}\frac{Tr(C^{L-1}Az(v)D(v))}{Tr(C^LA)}dv=1-\frac{1}{\alpha}\frac{Tr(C^{L-1}A)}{Tr(C^LA)}
\ee
in which we have used the algebra (\rf{alg}). Also for $\rho_{total}(L)$ we obtain the following expression using the algebra (\rf{alg}):
\be
\rho_{total}(L)=\int_{v_1}\frac{Tr(C^{L-1}z(v)D(v)A)}{Tr(C^LA)}dv=(\int_{v_1}\frac{vz(v)dv}{v+\beta-1})\frac{Tr(C^{L-1}A)}{Tr(C^LA)}.
\ee
Using (\rf{S}) and (\rf{R}), one finds $\rho_{total}(1)=\rho$ and also
$\rho_{total}(L)\big \vert_{L\rightarrow \infty}=\rho-(\alpha+\beta-1)\int_{v_1}\frac{\rho(v)}{v+\beta-1}dv$ in the phase (\rf{1sing1}).
Note that while the total density of the ordinary particles is constant directly in front of the impurity, it might be greater (if
$\alpha+\beta<1$), smaller (if $\alpha+\beta>1$) or equal (if $\alpha+\beta=1$) to $\rho$ immediately behind it. 
The same phenomenon might take place with a slight difference for the total density of the ordinary particles immediately
in front of the 
impurity in the phase (\rf{1sing2}) where we obtain $\rho_{total}(1)=\rho+\frac{(\alpha+\beta-1)(1-\rho)}{\alpha}$ 
and $\rho_{total}(L)\big \vert_{L\rightarrow \infty}=\rho$. In this case $\rho_{total}(1)$ can be greater (if
$\alpha+\beta>1$), smaller (if $\alpha+\beta<1$) or equall (if $\alpha+\beta=1$) to $\rho$. 
In order to see the behaviours of the total density profile of the ordinary particles on the ring, we have performed the Monte Carlo
simulation. We will discuss the obtained results in the next section. \\
In figure 2 we have plotted the phase diagram of 
our model for $\frac{d}{du}S\Big\vert_{u=v_1}<0$. The total density profile of the ordinary particles
(\rf{total}) obtained from the Monte Carlo simulation is shown as a curved line in each phase. 
One should note that $\rho_{total}(i)$ has two different behaviours in the phase (\rf{1sing1}). 
The same phenomenon can be seen in the phase (\rf{1sing2}) (see figure 2). 
As we have already mentioned, the line $\alpha+\beta=1$ is where the impurity behaves like the slowest ordinary particle. 
Simulation results show that the total density profile of the ordinary particles on this line is flat. 
These results are in agreement with the picture that one can obtain from our analytical results for the values of the total density 
of the ordinary particles immediately behind as well as in front of the impurity. 
\begin{figure}
\centering
\includegraphics[height=6cm,angle=0]{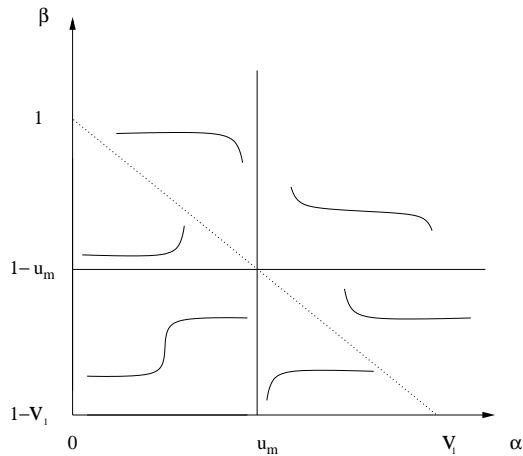}
\caption{\small{Phase diagram of the model where $\frac{d}{du}S\Big\vert_{u=v_1}<0$. The short lines show the behaviour of the total density profile
of the ordinary particles obtained from simulation results. The dotted line is $\alpha+\beta=1$.}}
\end{figure} 
In the region specified by (\rf{1sing4}) the single impurity provokes a macroscopic shock in the system.
The density profile of the ordinary particles in the shock phase, obtained from the Monte Carlo simulation, can be seen in figure 4. 
Nevertheless, when (\rf{cond}) becomes positive, the function (\rf{S}) is a monotonically increasing function of $u$ and the formal 
series (\rf{gener}) has only one singularity at $s_\alpha$:
\be
\la{2sing1}
s_\alpha=S({\bf z};\alpha) \ \ \mbox{for} \ \ \alpha<v_1.
\ee
This singularity comes from the disappearance of $1-\alpha B({\bf z};s_{\alpha})$ in the numerator of (\rf{gener2}).
The simulation data predicts that in this case, the total density profile of the ordinary particles (\rf{total}) has two different 
behaviours in the parameters space. Again, on the line $\alpha+\beta=1$ the total density profile in the steady state is flat.
In figure 3 we have plotted the phase diagram of our model for $\frac{d}{du}S\Big\vert_{u=v_1}>0$. \\
The physical quantities $V$ and $J_{total}$, which are given by (\rf{V}) and (\rf{J}) respectively, can be written in terms of 
the convergence radius of (\rf{gener}):
\bea
\begin{array}{l}
\la{vj}
V=(1-\int_{v_1}v z(v) dv)R({\bf z}) \\
J_{total}=\rho V + (R({\bf z})-V)
\end{array}
\eea 
\begin{figure}
\centering
\includegraphics[height=6cm,angle=0]{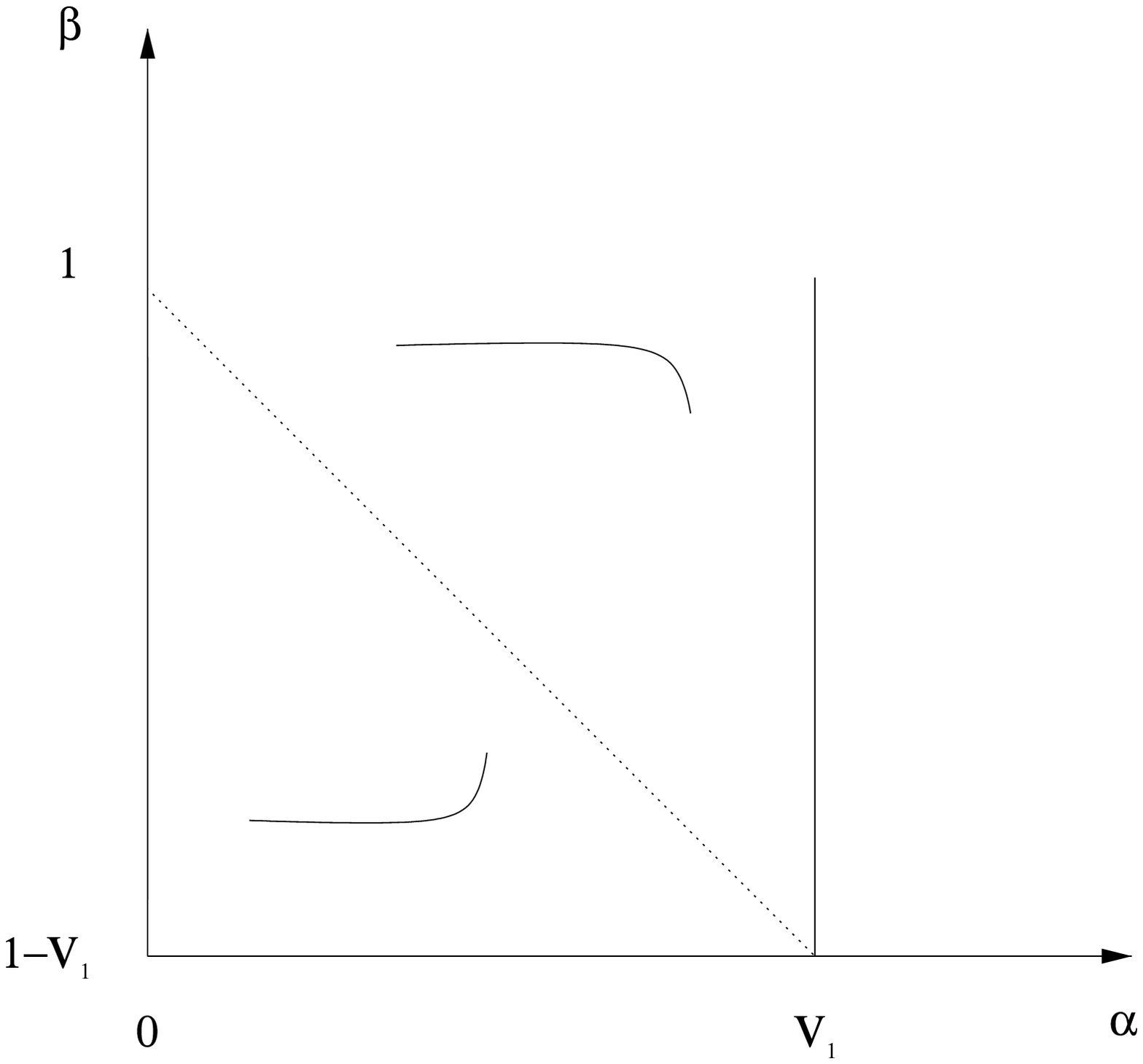}
\caption{\small{Phase diagram of the model where $\frac{d}{du}S\Big\vert_{u=v_1}>0$. The short curved lines show the bahaviour of the 
total density profile of the ordinary particles obtained from simulation results. The dotted line is $\alpha+\beta=1$.}}
\end{figure} 
By using (\rf{mean2}) and (\rf{vj}) one obtains the following expressions for $V$ and $J_{total}$ in each region of 
(\rf{1sing1})-(\rf{1sing4}):
\bea
\begin{array}{ll}
\la{vj2}
V=\alpha-\int_{v_1}v\rho(v)dv \ , \ J_{total}=(1-\rho)\int_{v_1}v\rho(v)dv  & \mbox{ for } \ \  \alpha<u_m,\beta>1-u_m\\
V=1-\beta-\int_{v_1}v\rho(v)dv \ , \ J_{total}=(1-\rho)\int_{v_1}v\rho(v)dv & \mbox{ for } \ \  \alpha>u_m,\beta<1-u_m\\
V=u_m-\int_{v_1}v\rho(v)dv \ , \ J_{total}=(1-\rho)\int_{v_1}v\rho(v)dv     & \mbox{ for } \ \  \alpha>u_m,\beta>1-u_m\\ 
V=(\alpha-\beta+1-u_m)-\int_{v_1}v\rho(v)dv \ , \ J_{total}=(\frac{\alpha(1-\beta)}{u_m}-V)(1-\rho)& \mbox{ for } \ \  
\alpha<u_m,\beta<1-u_m\\
\end{array}
\eea
As can be seen from (\rf{vj2}) both the speed of impurity and total current of the ordinary particles are continuous functions
of $\alpha$ and $\beta$. We should note that using the grand canonical ensemble procedure, one cannot find the exact expressions
of $V$ and $J_{total}$ in the shock phase except on the boundaries of this phase. This phenomenon has already been
observed in \cite{farhad,sasa2};  
nevertheless, the Monte Carlo simulation data in this phase predict that for a fixed value of $\alpha$ ($\beta$) these quantities 
are linear functions of $\beta$ ($\alpha$). Keeping in mind that these quantities change continuously over the 
boundaries, it is not difficult to conjecture the expressions given in the shock phase (\rf{vj2}); however, Monte Carlo results
confirm the correctness of these expressions. \\
Using (\rf{2sing1}), (\rf{mean2}) and (\rf{vj}), we also obtain the following expressions in the second regime:
\be
\la{vj3}
V=\alpha-\int_{v_1}v\rho(v)dv \ , \ J_{total}=(1-\rho)\int_{v_1}v\rho(v)dv \ \ \ \ \mbox{ for } \ \  \alpha<v_1\\
\ee

{\bf Remark:}\\
If we had released the condition $\alpha<v_1$ (the fact that the impurity should be the slowest particle on the ring), 
there would have found a change in the phase diagram of our model in (\rf{2sing1}). It can be shown that in this case 
an extra phase exists for $\alpha>v_1$ which is produced by the singularity of (\rf{S}) at $v_1$ \footnote{In fact (\rf{S}) has a
branch cut on a segment of the real axis begining from $u=v_1$ to $+\infty$.}:
\be
\la{2sing2}
s_{v_1}=S({\bf z};v_1) \ \ \mbox{for} \ \ \alpha>v_1.
\ee 
The speed of the impurity as well as the total current of the ordinary particles can also be obtained in this phase:  
\be
\la{vj4}
V=v_1-\int_{v_1}v\rho(v)dv \ , \ J_{total}=(1-\rho)\int_{v_1}v\rho(v)dv \ \ \ \ \mbox{ for } \ \  \alpha>v_1.\\
\ee  
\se{Special Cases and Simulation Results}
In this section we will check the validity of our analytical results obtained in the previous section by considering two 
different examples. As the first example we consider $\rho(v)=\rho\delta(v-1)$. 
In this case all the ordinary particles have the same hopping 
rates equal to the unity, and $\rho$ is their total density. This model has already been studied in \cite{m} as a simple exclusion
model for the study of shocks in one-dimension. The function (\rf{S}) has a very simple form in this case:
\be
S({\bf z};u)=\frac{u(1-u)}{1+(z-1)u}.
\ee
One can easily see that (\rf{cond}) is equal to $-\infty$ and also (\rf{bm2}) gives $u_m=1-\rho$; therefore,
the mean values of the speed of impurity and total current of particles should be obtained from (\rf{vj2}):
\bea
\begin{array}{ll}
\la{mal}
V=\alpha-\rho \ , \ J_{total}=\rho(1-\rho)  & \mbox{ for } \ \  \alpha<1-\rho,\beta>\rho\\
V=1-\beta-\rho \ , \ J_{total}=\rho(1-\rho)& \mbox{ for } \ \  \alpha>1-\rho,\beta<\rho\\
V=1-2-\rho \ , \ J_{total}=\rho(1-\rho)     & \mbox{ for } \ \  \alpha>1-\rho,\beta>\rho\\ 
V=\alpha-\beta \ , \ J_{total}=\rho(\alpha-\beta)+\beta(1-\alpha) & \mbox{ for } \ \  \alpha<1-\rho,\beta<\rho\\
\end{array}
\eea
The results (\rf{mal}) agree with those obtained in \cite{m}; however, the analytical calculations in \cite{m} show that the density profile 
of the ordinary particles in the first and the second phase of (\rf{mal}) has an exponential behaviour with two different characteristic 
length scales for $\alpha+\beta>1$ and $\alpha+\beta<1$. It is also known that the fourth region in (\rf{mal}) is where the density
profile of the ordinary particles has a shock structure.\\ 
As the second example, we consider the following density distribution function:
\be
\la{dist}
\rho(v)=\frac{\rho(n+1)}{(1-v_1)^{n+1}}(v-v_1)^n \ \ , \ \ n\geq0
\ee
in which $n$ is not necessarily an integer. This kind of distribution has also been considered in related contexts \cite{ev,k3}.
Using (\rf{mean2}) and (\rf{dist}) the condition (\rf{cond}) can be written as:
\be
\la{cond2}
\frac{d}{du}S\Big\vert_{u=v_1}\propto\frac{1-\rho}{v_1}-\frac{\rho(n+1)}{n(1-v_1)}.
\ee 
For $n>\frac{\rho v_1}{1-\rho-v_1}$ the sign of (\rf{cond2}) becomes positive and the phase diagram
should be obtained from (\rf{2sing1}). Taking $n=1$ and $\rho=0.6$ we will obtain the total density 
profile of the ordinary particles in each phase using the Monte Carlo simulation for $v_1=0.6$ and
$v_1=0.2$. The values of $u_m$ for $v_1>0.25$ should be obtained from (\rf{bm2}). 
For $\rho=0.6$ and $v_1=0.6$, using (\rf{bm2}) and (\rf{dist}), we obtain $u_m=0.338817$.
In figures 4 and 5 we have plotted the total density profile of the ordinary particles $\rho_{total}(i)$ obtained from the Monte 
Carlo simulation (as seen by the impurity) on a chain of length $L=500$ for $v_1=0.6$ at $\alpha=0.2$ and $\alpha=0.5$ respectively. 
Let us first explain the case $\alpha=0.2$. Depending on the value of $\beta$, 
three different phases can be seen: a shock phase, where the total density of the ordinary particles has a shock structure, and
two other phases in which the total density of the ordinary particles in the middle of the chain is equal to $\rho$.
It is also seen that the impurity produces a small disturbance just behind itself in these two phases.
\begin{figure}
\centering
\includegraphics[height=5cm,angle=0]{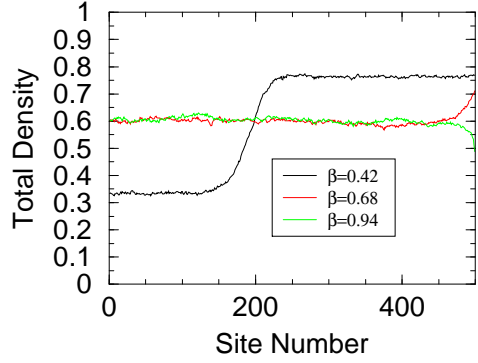}
\caption{\small{Plot of the total density profile of the ordinary particles on a ring of
length $L$=500 for $\alpha$=0.2, $v_1$=0.6, $\rho=0.6$ and different values of $\beta$.}}
\end{figure}
\begin{figure}
\centering
\includegraphics[height=5cm,angle=0]{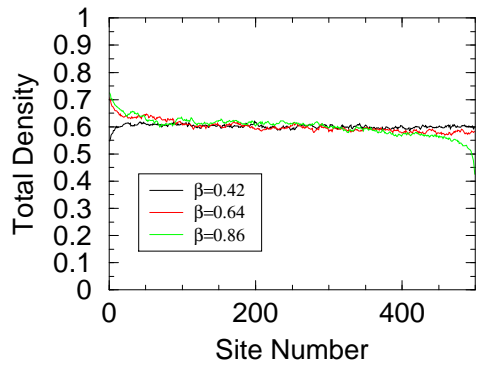}
\caption{\small{Plot of the total density profile of the ordinary particles on a ring of
length $L$=500 for $\alpha=0.5$, $v_1=0.6$, $\rho=0.6$ and different values of $\beta$.}}
\end{figure}
We have plotted $\rho_{total}(i)$ for $\alpha=0.5$ in figure 5. It can be seen that the total density profile of the
ordinary particles has again three different behaviours depending on the values of $\beta$. Simulation data also show that 
this behaviour is generic, at least for the the family of distribution functions given by (\rf{dist}).   
In figure 6 the speed impurity obtained from the Monte Carlo simulation is plotted as a function of $\beta$ for $\alpha=0.2$ and 
$\alpha=0.5$. As can be observed the speed of impurity is a decreasing function of $\beta$ up to a critical point $\beta_c$.
Analytical calculations predict $\beta_c=1-u_m=0.6611834$. Above the critical point the speed of the impurity is a constant
independent of $\beta$. This agrees with our previous results (\ref{vj2}). 
\begin{figure}
\centering
\includegraphics[height=6cm,angle=270]{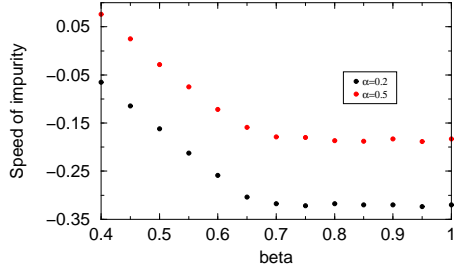}
\caption{\small{Plot of the speed of the impurity as a function of $\beta$ obtained from Monte Carlo simulation on a ring of
length $L$=500 for $v_1=0.6$, $\alpha=0.2$, $\alpha=0.5$ and $\rho=0.6$. 
The phase transition takes place at $\beta_c \simeq 0.66$}}
\end{figure}
As we mentioned above, the speed of impurity, as well as the total current of the ordinary particles in the 
shock phase are linear functions of $\alpha$ or $\beta$. In order to see this, we have plotted $J_{total}$ as a function of
$\beta$ for $\alpha=0.2$ obtained from the Monte Carlo simulation and compared with exact result (\rf{vj3}) in figure 7. 
As can be seen for a fixed value of $\alpha$, $J_{total}$ is a linear function of $\beta$ for $\beta < \beta_c \simeq 0.66$ 
while it remains constant above this point. 
\begin{figure}
\centering
\includegraphics[height=6cm,angle=270]{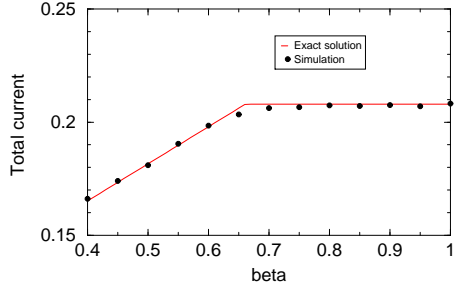}
\caption{\small{Plot of the total current of the ordinary particles as a function of $\beta$ obtained from the Monte Carlo simulation 
and also exact results on a ring of length $L$=500 for $v_1=0.6$, $\alpha=0.2$ and $\rho=0.6$. 
The phase transition takes place at $\beta_c \simeq 0.66$.}}
\end{figure}
Now for $v_1<0.25$ we expect that the phase diagram has only two different phases. We take $v_1=0.2$ so $\beta$ is limited to 
$\beta>1-v_1=0.8$. Figure 8 shows the total density profile of the ordinary particles for $v_1=0.2$, $\beta=0.85$ and two 
different values of $\alpha$. For a fixed value of $\beta$, one can see at most two different phases depending on the value of 
$\alpha$. 
In figure 9 we have plotted the speed of the impurity as a function of $\alpha$ for $\beta=0.8,0.85$ and $\beta=0.9$. 
Analytical results (\rf{vj3}) predict that for any values of $\beta$ the speed of impurity is an increasing function 
of $\alpha$. As can be seen in figure 9 the speed of impurity is independent of $\beta$ and also a linear function of 
$\alpha$ for $\alpha<0.2$. Again this is in agreement with our analytical results (\rf{vj3}).  
\begin{figure}
\centering
\includegraphics[height=5cm,angle=0]{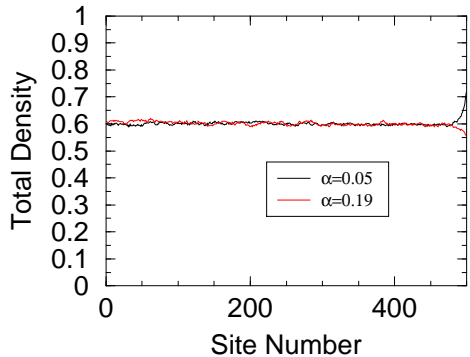}
\caption{\small{Plot of the total density profile of the ordinary particles on a ring of
length $L$=500 for $\beta$=0.85, $v_1$=0.2, $\rho=0.6$ and different values of $\alpha$.}}
\end{figure}
\begin{figure}
\centering
\includegraphics[height=6cm,angle=270]{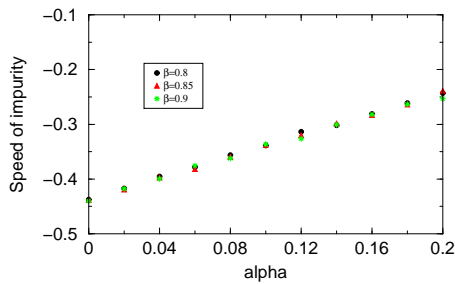}
\caption{\small{Plot of the speed of the impurity as a function of $\alpha$ obtained from the Monte Carlo simulation on a ring of
length $L$=500 for $\beta=0.8,0.85,0.9$, $v_1=0.2$ and, $\rho=0.6$. 
It is seen that the speed of impurity is independent of $\beta$.}}
\end{figure}

{\bf Remark:}\\
As we mentioned above, the analytical results for the case $\frac{d}{du}S\Big\vert_{u=v_1}>0$
suggest that the phase diagram in the case $0<\alpha<+\infty$ is made up of two different 
phases given by (\rf{2sing1}) and (\rf{2sing2}). The Monte Carlo simulations for this case have also been carried out. 
Our results confirm both the phase transition point $\alpha_c=v_1$
and also the behaviours of $V$ and $J_{total}$ obtained analytically. In figure 10 we have plotted the speed of impurity as a function of
$\alpha$ for $\beta=0.8,0.85$ and $\beta=0.9$. As can be seen the speed of impurity is an increasing function of $\alpha$ up to the point 
$\alpha \simeq 0.2$ which is exactly the transition point in this case. Above the transition point the speed of impurity $V$ is no longer
a function of $\alpha$, as the analytical results had predicted. 
\begin{figure}
\centering
\includegraphics[height=6cm,angle=270]{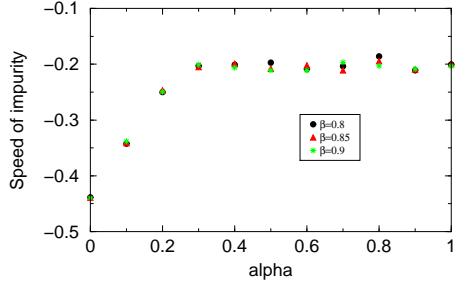}
\caption{\small{Plot of the speed of the impurity as a function of $\alpha$ for $0 < \alpha < 1$. The results are
obtained from the Monte Carlo simulation on a ring of length $L$=500 for $\beta=0.8,0.85,0.9$, $v_1=0.2$ and $\rho=0.6$. 
It is seen that the speed of impurity is independent of $\beta$ for $\alpha>\alpha_c\simeq 0.2$.}}
\end{figure}
The density profile of the ordinary particles on the ring $\rho_{total}(i)$ in the second phase ($\alpha>v_1$) is the same as
the phase (\rf{1sing3}). 
\se{Conclusion}
In this paper we considered the multi-species ASEP (first introduced in \cite{k1}) with periodic boundary condition in the presence of a 
single impurity. Using the MPA, we studied the stationary state properties 
of this model by introducing a grand canonical ensemble. We obtained not only the exact phase structure
but also exact expressions for the mean speed of the impurity $V$ and the total current of the ordinary particles $J_{total}$  
in the thermodynamic limit. The Monte Carlo Simulation data confirms the correctness of our analytical calculations at least for a large
family of the density distribution functions given by (\rf{dist}). \\
Depending on the sign of (\rf{cond}), which is only a function of the density distribution 
of the ordinary particles on the ring, the phase diagram of the model can exist in two different regimes.
Both the analytical and the Monte Carlo simulation data show that, as far as the sign of (\rf{cond}) is negative, 
the phase diagram of our model has nearly the same structre of the phase diagram of the one-species model \cite{m}. 
However, there are some differences: in comparison to the one-species ASEP
the speed of impurity in the shock phase (\rf{1sing4}), which is in fact the speed of the shock in this phase, does not have the 
simple form $\alpha-\beta$; 
instead it depends on the density distribution function of the ordinary particles on the ring. Also
in the multi-species case, the boundaries of the phase diagram in the case $\frac{d}{du}S\Big\vert_{u=v_1}<0$ are specified by the 
quantity $u_m$ which is the solution of (\rf{bm2}) and its value depends on $\rho(v)$. 
In the special limit $\rho(v)=v\delta(v-1)$ all our results converge on those obtained in \cite{m}. 
All of the analytical results as well as the Monte Carlo simulation data confirm that
if the sign of (\rf{cond}) is positive, the phase diagram has no longer four different phases; instead it has one phase.
The impurity, by definition, should be the slowest particle in the ensemble of the particles on the ring; however, if
one releases this condition then a new phase appears. The phase transition point $\alpha_c=v_1$ can be calculated analytically 
and also confirmed by simulation. As we mentioned, this new regime cannot be seen in the one-species ASEP.
\se{Acknowledgements}
I would like to thank V. Karimipour for critical reading of the manuscript and also for his valuable comments. 
I also thank M. E. Fouladvand for his comments. I appreciate R. W. Sorfleet for his help during preparation of this paper.\\
{\large {\bf References}}


\begin{enumerate}
\bibitem{sp} F. Spitzer, Adv. Math. {\bf 5},246(1970).
\bibitem{sz} B. Schmittmann and R. K. P. Zia in "{\it Phase transitions
and critical phenomena}" vol. 17, eds. C. Domb and J. Lebowitz (London,
Academic Press, 1995).
\bibitem{l} T. M. Ligget, {\it Interacting Particle Systems} (Springer-Verlag, New
York, 1985).
\bibitem{sph} H. Spohn, {\it Large Scale Dynamics of Interacting
Particles} (Springer-Verlag, New York, 1991).
\bibitem{kr1} J. Krug and H. Spohn, Phys. Rev. {\bf A38}, 4271(1988).
\bibitem{kd} D. Kandel and E. Domany, J. Stat. Phys {\bf 58}, 685(1992).
\bibitem{kr2} J. Krug and H. Spohn, Kinetic roughening of growing surfaces in {\it Solids Far From Equilibrium},
C. Godr\'eche, ed. (Cambridge University Press,1991).
\bibitem{chsa} D. Chowdhury, L. Santen and A. Schadschneider, Phys. Rep. {\bf 329}, 199(2000).
\bibitem{mac} J. T. MacDonald, J. H. Gibbs and A. C. Pipkin, Biopolymers {\bf 6}, 1(1968).
\bibitem{schr} G. M. Sch\"utz; Integrable stochastic processes in {\it Phase
Transitions and Critical Phenomena}; eds. C. Domb and J. Lebowitz (Academic Press, New York, 1999).    
\bibitem{d} B. Derrida, Phys. Rep. {\bf 301}, 65(1998).
\bibitem{dehp} B. Derrida, M.R. Evans, V.Hakim and V. Pasquier,
J. Phys. {\bf A26}, 1493(1993).
\bibitem{de} B. Derrida and M. R. Evans in {\it " Non-Equilibrium
Statistical Mechanics in
one Dimension}", V. Privman ed. (Cambridge University Press, 1997).
\bibitem{sand} S. Sandow, Phys. Rev. {\bf E50}, 2660(1994). 
\bibitem{esri} F. H. L. Essler and V. Rittenberg, J. Phys. {\bf A29}, 3375(1996).
\bibitem{sas1} T. Sasamoto, J. Phys. {\bf A32}, 7109(1999). 
\bibitem{djls} B. Derrida, S. A. Janowsky, J. L. Lebowitz and E. R. Speer,
J. Stat. Phys. {\bf 78}, 813(1993).
\bibitem{m} K. Mallick, J. Phys. {\bf A29}, 5375(1996).
\bibitem{farhad} F. H. Jafarpour, J. Phys. {\bf A33}, 1797(2000).  
\bibitem{sasa2} T. Sasamoto, cond-mat/9910483.
\bibitem{lpk} H. W. Lee, V. Popkov and D. Kim, J. Phys. {\bf A30}, 8497(1997).
\bibitem{k1} V. Karimipour, Phys. Rev. {\bf E59}, 205(1999).
\bibitem{k2} V. Karimipour, Europhys. Letts. {\bf 47}(3), 304(1999).
\bibitem{k3} M. Khorrami and V. Karimipour, J. Stat. Phys. {\bf 100}, 999(2000).
\bibitem{ks} K. Krebs and S. Sandow; J. Phys. {\bf A30}, 3165(1997); K. Krebs, preprint cond-mat/9910452.
\bibitem{efgm} M. R. Evans, D. P. Foster, C. Godr\'eche and D. Mukamel, J. Stat. Phys. {\bf 80}, 69(1995). 
\bibitem{ev} M. R. Evans, J. Phys. {\bf A30}, 5669(1997);
Europhys. Lett. {\bf 36}, 13(1996).
\end{enumerate}
\end{document}